\documentstyle[aps,epsf]{revtex}
\draft
\begin{document}

\title{Supernova Neutrinos and the Neutrino Masses}
\author{J.~F. Beacom}
\address{Department of Physics, California Institute of Technology\\
         Pasadena, CA 91125, USA}
\date{January 2, 1999}
\maketitle

\begin{abstract}

Core-collapse supernovae emit of order $10^{58}$ neutrinos and
antineutrinos of all flavors over several seconds, with average
energies of 10--25 MeV.  In the Sudbury Neutrino Observatory (SNO), a
future Galactic supernova at a distance of 10 kpc would cause several
hundred events.  The $\nu_\mu$ and $\nu_\tau$ neutrinos and
antineutrinos are of particular interest, as a test of the supernova
mechanism.  In addition, it is possible to measure or limit their
masses by their delay (determined from neutral-current events)
relative to the $\bar{\nu}_e$ neutrinos (determined from
charged-current events).  Numerical results are presented for such a
future supernova as seen in SNO.  Under reasonable assumptions, and in
the presence of the expected counting statistics, a $\nu_\mu$ or
$\nu_\tau$ mass down to about 30 eV can be simply and robustly
determined.  This seems to be the best technique for direct
measurement of these masses.

\end{abstract}

\pacs{14.60.Pq, 97.60.Bw, 25.30.Pt}


\section{Introduction}

Whether or not neutrinos have mass and are mixed are subjects of great
current interest.  Many experiments are now or will soon be searching
for neutrino flavor mixing in a variety of circumstances.  The
strongest evidence for mixing (which implies mass) so far comes from
the atmospheric neutrino experiments.  However, all of these
experiments by their nature are only sensitive to the differences of
neutrino masses, and not the mass scale.

The absolute scale of the neutrino masses is an important probe of
physics beyond the standard model of particle physics.  In addition,
if the neutrinos have masses of order a few eV or more, they may be an
important part of the dark matter in the universe.  Direct mass
measurements from decay kinematics do not place very stringent limits:
$m_{\nu_e} \lesssim 5$ eV\cite{Belesev}, $m_{\nu_\mu} < 170$
keV\cite{RPP}, and $m_{\nu_\tau} < 18$ MeV\cite{RPP}.  It will be very
difficult to significantly improve these limits with terrestrial
experiments.

A core-collapse supernova is a tremendous source of neutrinos and
antineutrinos of all flavors.  Since the current limit on the $\nu_e$
mass is comparatively low, the $\nu_\mu$ and $\nu_\tau$ masses could
be measured by their time-of-flight delay relative to the $\nu_e$ and
$\bar{\nu}_e$.  Since they have energies only of order 25 MeV, the
$\nu_\mu$ and $\nu_\tau$ can be detected only by their neutral-current
interactions.  While they also have neutral-current interactions, the
$\nu_e$ and $\bar{\nu}_e$ will be detected primarily by their
charged-current interactions.

Even a tiny mass will make the velocity slightly less than for a
massless neutrino, and over the large distance to a supernova will
cause a measurable delay in the arrival time.  A neutrino with a mass
$m$ (in eV) and energy $E$ (in MeV) will experience an
energy-dependent delay (in s) relative to a massless neutrino in
traveling over a distance D (in 10 kpc, approximately the distance
to the Galactic center) of
\begin{equation}
\Delta t(E) = 0.515
\left(\frac{m}{E}\right)^2 D\,,
\label{eq:delay}
\end{equation}
where only the lowest order in the small mass has been kept.

If the neutrino mass is nonzero, lower-energy neutrinos will arrive
later, leading to a correlation between neutrino energy and arrival
time.  Using this idea, Ref.~\cite{Totani} has shown that the next
supernova will allow sensitivity to a $\nu_e$ mass down to about 3 eV,
comparable to the terrestrial limit.  Since the neutrino energy is
{\it not} measured in neutral-current interactions a similar technique
cannot be used for the $\nu_\tau$ mass.  (The incoming neutrino energy
is not determined since a complete kinematic reconstruction of the
reaction products is typically not possible.)

Instead, the strategy for measuring the $\nu_\tau$ mass is to look at
the difference in time-of-flight between the neutral-current events
(mostly $\nu_\mu$,$\nu_\tau$,$\bar{\nu}_\mu$, and $\bar{\nu}_\tau$)
and the charged-current events (just $\nu_e$ and $\bar{\nu}_e$).  We
assume that the $\nu_\mu$ is massless and will ask what limit can be
placed on the $\nu_\tau$ mass.  There are three major complications to
a simple application of Eq.~(\ref{eq:delay}): (i) The neutrino
energies are not fixed, but are characterized by spectra; (ii) The
neutrino pulse has a long intrinsic duration of about 10 s, as
observed for SN1987A; and (iii) The statistics are finite.

One possible neutral-current signal is the excitation of $^{16}{\rm
O}$, followed by detectable gamma emission~\cite{LVK}.  In
SuperKamiokande (SK), which has a target volume of 32 kton of light
water, this would cause about 710 events.  This signal would allow
sensitivity to a $\nu_\tau$ mass as low as about 45 eV~\cite{SKpaper}.

Another possible neutral-current signal is deuteron breakup, followed
by neutron detection.  In the Sudbury Neutrino Observatory (SNO),
which has a target volume of 1 kton of heavy water, this would cause
about 485 events.  (This detector also has a light-water target with
an active volume of about 1.4 kton).  While the statistics are
somewhat lower than for SK, the energy dependence of the cross section
is less steep and emphasizes lower energies and hence longer delays,
leading to a sensitivity to a $\nu_\tau$ mass down to about 30
eV~\cite{SNOpaper}

Since one expects a type-II supernova about every 30 years in our
Galaxy, there is a good chance that these mass limits can be
dramatically improved in the near future.


\section{Production and detection of supernova neutrinos}

When the core of a large star ($M \ge 8 M_{\odot}$) runs out of
nuclear fuel, it collapses to proto-neutron star.  About 99\% of the
gravitational binding energy change, about $3 \times 10^{53}$ ergs, is
carried away by neutrinos.  Because of the high density, they diffuse
outward over a timescale of several seconds.  When they are within
about one mean free path of the edge, they escape freely, with a
thermal spectrum (approximately Fermi-Dirac) characteristic of the
surface of last scattering.  Because different flavors have different
interactions with the matter, the temperatures are different.  The
$\nu_\mu$ and $\nu_\tau$ neutrinos and their antiparticles have a
temperature of about 8 MeV (or $\langle E \rangle \simeq$ 25 MeV).
The $\bar{\nu}_e$ neutrinos have a temperature of about 5 MeV
($\langle E \rangle \simeq$ 16 MeV), and the $\nu_e$ neutrinos have a
temperature of about 3.5 MeV ($\langle E \rangle \simeq$ 11 MeV).  The
luminosities of the different neutrino flavors are approximately equal
at all times.  The neutrino luminosity rises quickly over a time of
order 0.1 s, and then falls over a time of order several seconds,
roughly like an exponential with a time constant $\tau$ = 3 s.  The
detailed form of the neutrino luminosity used below is less important
than the general shape features and their characteristic durations.

For thermal spectra which are constant in time, and for equal
luminosities among the different flavors, the scattering rate for
a given reaction can be written as:
\begin{equation}
\frac{dN_{sc}}{dt} = C
\int dE\,f(E) \left[\frac{\sigma(E)}{10^{-42} {\rm cm}^2}\right]
\left[\frac{L(t - \Delta t(E))}{E_B/6}\right]\,,
\label{eq:rate}
\end{equation}
where $f(E)$ is the neutrino energy spectrum, $\sigma(E)$ the cross
section, and $L(t)$ the luminosity.  For a massless neutrino (i.e.,
the charged-current events), $\Delta t(E) = 0$, and the time
dependence of the scattering rate is simply the time dependence of the
luminosity.  For a massive neutrino (i.e., the neutral-current
$\nu_\tau$ events), the time dependence of the scattering rate is
additionally dependent on the mass effects, as written.  The overall
constant is
\begin{equation}
C = 8.28
\left[\frac{E_B}{10^{53} {\rm\ ergs}}\right]
\left[\frac{1 {\rm\ MeV}}{T}\right]
\left[\frac{10 {\rm\ kpc}}{D}\right]^2
\left[\frac{{\rm det.\ mass}}{1 {\rm\ kton}}\right]
\,n\,,
\label{eq:C}
\end{equation}
where $E_B$ is the total binding energy release, $T$ is the spectrum
temperature, $D$ is the distance to the supernova, and $n$ is the
number of targets per molecule for the given reaction.  For a
light-water detector, the initial coefficient in $C$ is 9.21 instead
of 8.28.

The Sudbury Neutrino Observatory, though primarily intended for solar
neutrinos, also makes an excellent detector for supernova neutrinos.
Electrons and positrons will be detected by their \v{C}erenkov
radiation, and gammas via secondary electrons and positrons.  Neutrons
will be detected by one of three possible modes, depending on the
detector configuration.  The key neutral-current reaction is deuteron
breakup: $\nu + d \rightarrow \nu + p + n$ and $\bar{\nu} + d
\rightarrow \bar{\nu} + p + n$, with thresholds of 2.22 MeV.  Some
other relevant reactions are given in Table~I.


\section{Signature of a small neutrino mass}


\subsection{General description of the data}

As noted, for a massless neutrino ($\nu_e$ or $\bar{\nu}_e$) the time
dependence of the scattering rate is simply the time dependence of the
luminosity.  For a massive neutrino ($\nu_\tau$), the time dependence
of the scattering rate additionally depends on the delaying effects of
a mass.  To search for these effects, we define two rates: a Reference
$R(t)$ containing only massless events, and a Signal $S(t)$ containing
some fraction of massive events (along with some massless events which
cannot be separated).

The Reference $R(t)$ can be formed in various ways, for example from
the charged-current reaction $\bar{\nu}_e + p \rightarrow e^+ + n$ in
the light water of either SK or SNO (the former with a much better
precision).

The primary component of the Signal $S(t)$ in SNO is the 485
neutral-current events on deuterons.  With the hierarchy of
temperatures assumed here, these events are 18\% ($\nu_e +
\bar{\nu}_e$), 41\% ($\nu_\mu + \bar{\nu}_\mu$), and 41\% ($\nu_\tau +
\bar{\nu}_\tau$).  The flavors of the neutral-current events of course
cannot be distinguished.  Under our assumption that only $\nu_\tau$ is
massive, there is already some unavoidable dilution of $S(t)$.

In Figure~1, $S(t)$ is shown under different assumptions about the
$\nu_\tau$ mass.  The shape of $R(t)$ is exactly that of $S(t)$ when
$m_{\nu_\tau} = 0$, though the number of events in $R(t)$ will be
different.  The rates $R(t)$ and $S(t)$ will be measured with finite
statistics, so it is possible for statistical fluctuations to obscure
the effects of a mass when there is one, or to fake the effects when
there is not.  We determine the mass sensitivity in the presence of
the statistical fluctuations by Monte Carlo modeling.  We use the
Monte Carlo to generate representative statistical instances of the
theoretical forms of $R(t)$ and $S(t)$, so that each run represents
one supernova as seen in SNO.  The best test of a $\nu_\tau$ mass
seems to be a test of the average arrival time $\langle t \rangle$.
Any massive component in $S(t)$ will always increase $\langle t
\rangle$, up to statistical fluctuations.


\subsection{$\langle t \rangle$ analysis}

Given the Reference $R(t)$ (i.e., the charged-current events), the
average arrival time is defined as
\begin{equation}
\langle t \rangle_R = \frac{\sum_k t_k}{\sum_k 1}\,,
\end{equation}
where the sum is over events in the Reference.  The effect of the
finite number of counts $N_R$ in $R(t)$ is to give $\langle t \rangle_R$
a statistical error:
\begin{equation}
\delta\left(\langle t \rangle_R\right) =
\frac{\sqrt{\langle t^2 \rangle_R - \langle t \rangle^2_R}}{\sqrt{N_R}}\,.
\label{eq:tRerror}
\end{equation}
For a purely exponential luminosity,
$\langle t \rangle_R =
\sqrt{\langle t^2 \rangle_R - \langle t \rangle^2_R} = \tau$.

Given the Signal $S(t)$ (i.e., the neutral-current events), the
average arrival time $\langle t \rangle_S$ and its error
$\delta\left(\langle t \rangle_S\right)$ are defined similarly.  The
widths of $R(t)$ and $S(t)$ are similar, each of order $\tau = 3$ s
(the mass increases the width of $S(t)$ only slightly for small
masses.)  The signal of a mass is that the measured value of $\langle
t \rangle_S - \langle t \rangle_R$ is greater than zero with
statistical significance.

Using the Monte Carlo, we analyzed $10^4$ simulated supernova data
sets for a range of $\nu_\tau$ masses.  For each data set, $\langle t
\rangle_S - \langle t \rangle_R$ was calculated and its value
histogrammed.  These histograms are shown in the upper panel of Fig.~2
for a few representative masses.  (Note that the number of Monte Carlo
runs only affects how smoothly these histograms are filled out, and
not their width or placement.)  These distributions are characterized
by their central point and their width, using the 10\%, 50\%, and 90\%
confidence levels.  That is, for each mass we determined the values of
$\langle t \rangle_S - \langle t \rangle_R$ such that a given
percentage of the Monte Carlo runs yielded a value of $\langle t
\rangle_S - \langle t \rangle_R$ less than that value.  With these
three numbers, we can characterize the results of complete runs with
many masses much more compactly, as shown in the lower panel of
Fig.~2.  Given an experimentally determined value of $\langle t
\rangle_S - \langle t \rangle_R$, one can read off the range of masses
that would have been likely (at these confidence levels) to have given
such a value of $\langle t \rangle_S - \langle t \rangle_R$ in one
experiment.  From the lower panel of Fig.~2, we see that SNO is
sensitive to a $\nu_\tau$ mass down to about 30 eV if the SK $R(t)$ is
used, and down to about 35 eV if the SNO $R(t)$ is used.

We also investigated the dispersion of the event rate in time as a
measure of the mass.  A mass alone causes a delay, but a mass and an
energy spectrum also cause dispersion.  We defined the dispersion as
the change in the width $\sqrt{\langle t^2 \rangle_S - \langle t
\rangle^2_S} - \sqrt{\langle t^2 \rangle_R - \langle t \rangle^2_R}$.
We found that the dispersion was not statistically significant until
the mass was of order 80 eV or so; however, for such a large mass the
statistical significance of $\langle t \rangle_S - \langle t
\rangle_R$ cannot be missed.  This means that the average delay is
well-characterized by a single energy, which for SNO is $E_c \simeq
32$ MeV.


\section{Conclusions and discussion}

One of the key points of our technique is that the abundant
$\bar{\nu}_e$ events can be used to calibrate the neutrino luminosity
of the supernova and to define a clock by which to measure the delay
of the $\nu_\tau$ neutrinos.  The internal calibration substantially
reduces the model dependence of our results, and allows us to be
sensitive to rather small masses.  Our calculations indicate that a
significant delay can be seen for $m = 30$ eV with the SNO data,
corresponding to a delay in the average arrival time of about 0.15 s.
Even though the duration of the pulse is expected to be of order 10 s,
such a small average delay can be seen because several hundred events
are expected.  Without such a clock, one cannot determine a mass limit
with the $\langle t \rangle_S - \langle t \rangle_R$ technique
advocated here, since the absolute delay would be unknown.  Instead,
one would have to constrain the mass from the observed dispersion of
the events; only for a mass of $m = 150$ eV or greater would the pulse
become significantly broader than expected from theory.

Moreover, the technique used here allows accurate analytic estimates
of the results, so that it is easy to see how the conclusions would
change if different input parameters were used.  If the $\nu_\tau$
mass is very small, and a only a limit is placed, then this scales as
$m_{limit} \sim T^{3/4}\, \sqrt{\tau}$, where $T$ is the $\nu_\mu$ and
$\nu_\tau$ temperature and $\tau$ is the luminosity
timescale~\cite{SNOpaper}.  Thus the final result is relatively
insensitive to the supernova parameters in their expected ranges.
Additionally, this is {\it independent} of the distance $D$.  Because
of obscuration by dust, it may be difficult to observe the light from
a future Galactic supernova.  It is therefore rather important that
this does not affect the ability to place a limit on the $\nu_\tau$
mass.  In Ref.~\cite{SNpointing}, we have discussed how a supernova
could be located by its neutrinos, perhaps in advance or independently
of the light.

The observation of the neutrino signal of a future Galactic supernova
will be extremely significant test of the physics involved.  It will
allow, among other things, determination of the imprecisely-known
supernova neutrino emission parameters.  In addition, we hope to be
able to use the same data to determine or constrain neutrino
properties.  In Refs.~\cite{SKpaper,SNOpaper}, we discuss how both of
these goals can be achieved simultaneously, with or without the
additional complication of neutrino oscillations.

Despite the long intrinsic duration of the supernova neutrino pulse
and the spectra of neutrino energies, it is in fact possible to
discern even a small $\nu_\tau$ mass by a time-of-flight measurement.
The results are that SK and SNO are sensitive to a $\nu_\tau$ mass as
low as about 45 eV and 30 eV, respectively.  In the above, we
considered that the $\nu_\mu$ is massless and the $\nu_\tau$ is
massive.  Since they cannot be distinguished experimentally, the limit
in fact applies to both $\nu_\mu$ and $\nu_\tau$.  These results
include the effects of the finite statistics, and are relatively
insensitive to uncertainties in some of the key supernova parameters.
When the next Galactic supernova is observed, the $\nu_\tau$ mass
limit will be improved by nearly 6 orders of magnitude.  The
importance of this result is highlighted by its significance to both
cosmology and particle physics.  So that the universe is not
overclosed, the sum of the stable neutrino masses must be less than
about 100 eV.  Some seesaw models of the neutrino masses predict a
$\nu_\tau$ mass as large as about 30 eV~\cite{Bludman}).  As noted,
this seems to be the best technique for direct measurement of the
$\nu_\mu$ and $\nu_\tau$ masses.


\section*{ACKNOWLEDGMENTS}

I acknowledge support as a Sherman Fairchild fellowship from Caltech,
and I thank Petr Vogel for his collaboration on
Refs.~\cite{SKpaper,SNOpaper,SNpointing}.



\newpage

\begin{table}
\caption{Calculated numbers of events expected in SNO for a supernova
at 10 kpc.  In rows with two reactions listed, the number of
events is the total for both.  The notation $\nu$ indicates the sum
of $\nu_e$, $\nu_\mu$, and $\nu_\tau$, though they do not contribute
equally to a given reaction, and $X$ indicates either $n + ^{15}$O or
$p + ^{15}$N.}
\begin{tabular}{l|r}
\multicolumn{2}{c}{Events in 1 kton D$_2$O}\\
\hline
$\nu + d \rightarrow \nu + p + n$ & 485 \\
$\bar{\nu} + d \rightarrow \bar{\nu} + p + n$ & \\
\hline
$\nu_e + d \rightarrow e^- + p + p$ & 160 \\
$\bar{\nu}_e + d \rightarrow e^+ + n + n$ & \\
\hline
$\nu + ^{16}{\rm O} \rightarrow \nu + \gamma + X$ & 20 \\
$\bar{\nu} + ^{16}{\rm O} \rightarrow \bar{\nu} + \gamma + X$ & \\
\hline
$\nu + ^{16}{\rm O} \rightarrow \nu + n + ^{15}{\rm O}$ & 15 \\
$\bar{\nu} + ^{16}{\rm O} \rightarrow \bar{\nu} + n + ^{15}{\rm O}$ & \\
\hline
$\nu + e^- \rightarrow \nu + e^-$ & 10 \\
$\bar{\nu} + e^- \rightarrow \bar{\nu} + e^-$ & \\
\hline\hline
\multicolumn{2}{c}{Events in 1.4 kton H$_2$O}\\
\hline
$\bar{\nu}_e + p \rightarrow e^+ + n$ & 365 \\
\hline
$\nu + ^{16}{\rm O} \rightarrow \nu + \gamma + X$ & 30 \\
$\bar{\nu} + ^{16}{\rm O} \rightarrow \bar{\nu} + \gamma + X$ & \\
\hline
$\nu + e^- \rightarrow \nu + e^-$ & 15 \\
$\bar{\nu} + e^- \rightarrow \bar{\nu} + e^-$ & \\
\end{tabular}
\end{table}


\newpage
\centerline{\bf Figure Captions}

\bigskip

FIG. 1. The expected event rate for the Signal $S(t)$ at SNO in the
absence of fluctuations for different $\nu_\tau$ masses, as follows:
solid line, 0 eV; dashed lines, in order of decreasing height: 20, 40,
60, 80, 100 eV.  Of 535 total events, 100 are massless ($\nu_e +
\bar{\nu}_e$), 217.5 are massless ($\nu_\mu + \bar{\nu}_\mu$), and
217.5 are massive ($\nu_\tau + \bar{\nu}_\tau$).  These totals count
events at all times; in the figure, only those with $t \le 9$ s are
shown.

\bigskip

FIG. 2. The results of the $\langle t \rangle$ analysis for a massive
$\nu_\tau$, using the Signal $S(t)$ from SNO defined in the text.  In
the upper panel, the relative frequencies of various $\langle t
\rangle_S - \langle t \rangle_R$ values are shown for a few example
masses.  The solid line is for the results using the SK Reference
$R(t)$, and the dotted line for the results using the SNO $R(t)$.  In
the lower panel, the range of masses corresponding to a given $\langle
t \rangle_S - \langle t \rangle_R$ is shown.  The dashed line is the
50\% confidence level.  The upper and lower solid lines are the 10\%
and 90\% confidence levels, respectively, for the results with the SK
$R(t)$.  The dotted lines are the same for the results with the SNO
$R(t)$.


\newpage

\begin{figure}
\epsfxsize=6.5in \epsfbox{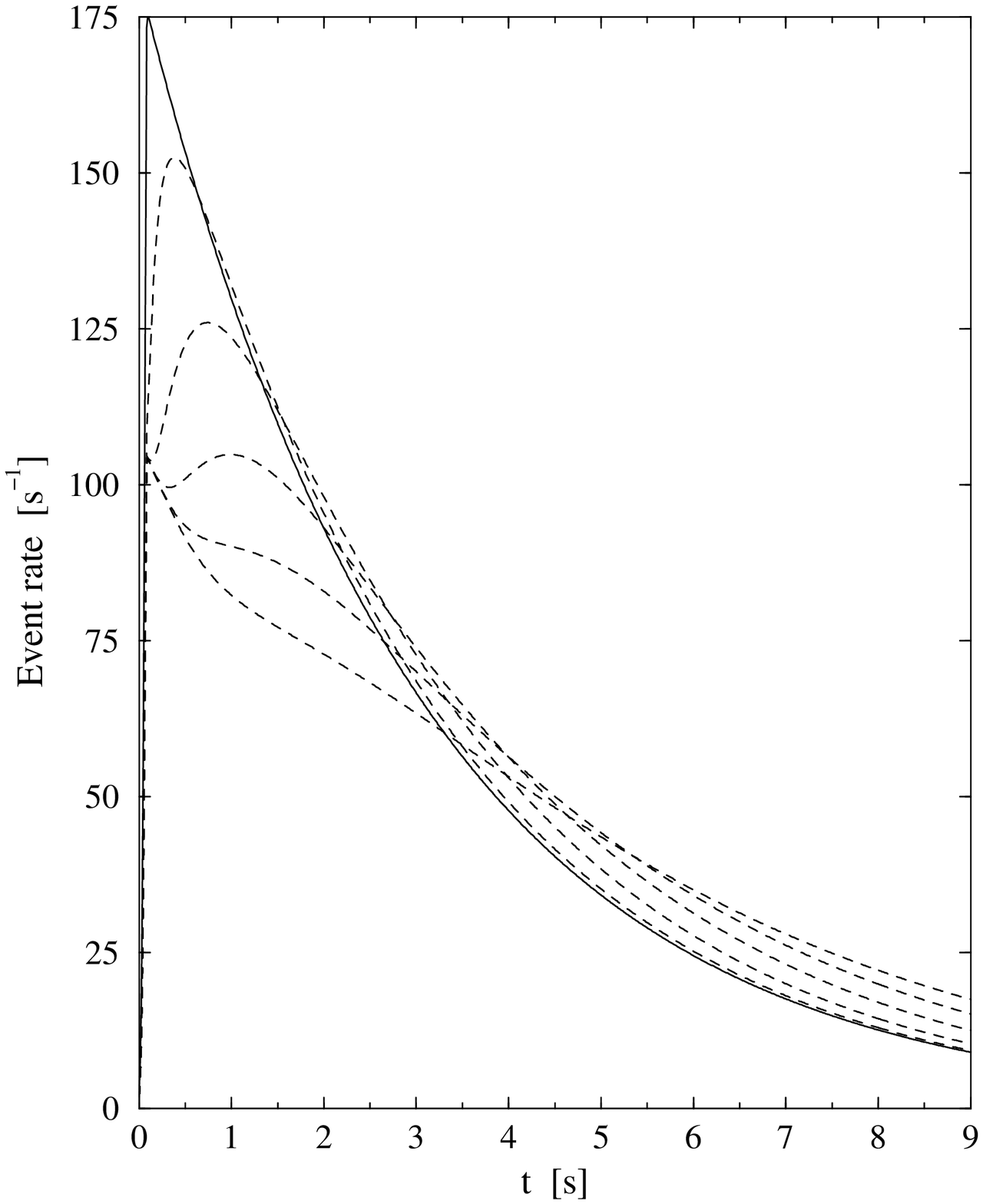}
\bigskip \centerline{{\large Figure 1}}
\end{figure}

\begin{figure}
\epsfxsize=6.5in \epsfbox{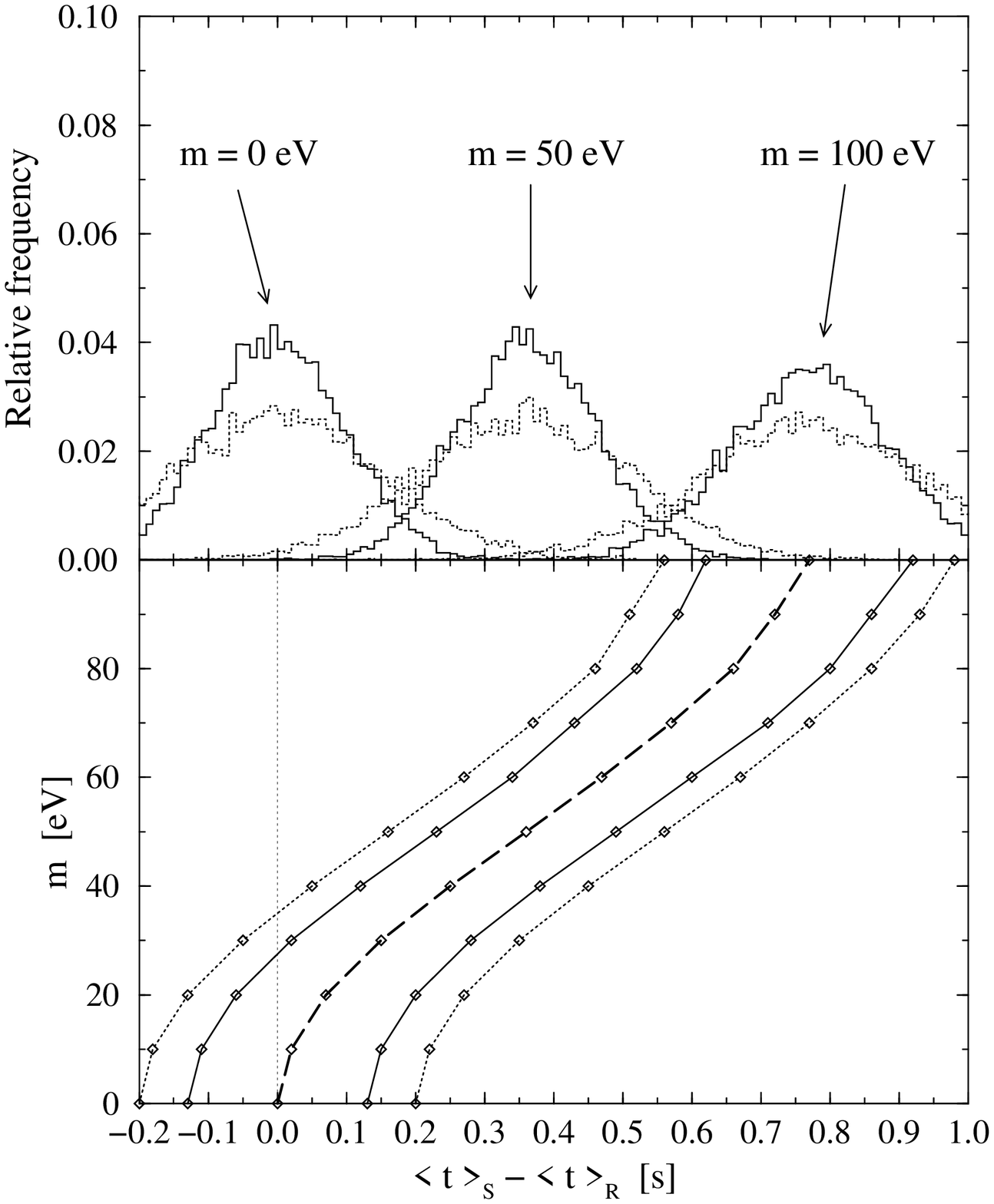}
\bigskip \centerline{{\large Figure 2}}
\end{figure}


\end{document}